\documentclass[10pt,english]{elsarticle}
\usepackage[utf8]{inputenc}
\usepackage[a4paper]{geometry}
\usepackage{amsmath}
\usepackage{amssymb}
\usepackage{setspace}
\usepackage{esint}
\usepackage{babel}

\begin{document}

\title{A relation between deformed superspace and Lee-Wick higher-derivative
theories}

\author{M. Dias}
\ead{marco.dias@unifesp.br}
\address{Universidade Federal de São Paulo, Departamento de Ciências Exatas
e da Terra, Rua Prof. Artur Riedel, 275, 09972-270, Diadema, SP, Brazil}

\author{A. F. Ferrari}
\ead{alysson.ferrari@ufabc.edu.br}
\address{Universidade Federal do ABC, Centro de Ciências Naturais e Humanas,
Rua Santa Adélia, 166, 09210-170, Santo André, SP, Brazil.}

\author{C. A. Palechor}
\ead{caanpaip@gmail.com}
\address{Universidade Federal do ABC, Centro de Ciências Naturais e Humanas,
Rua Santa Adélia, 166, 09210-170, Santo André, SP, Brazil.}

\author{C. R. Senise Jr.}
\ead{carlos.senise@unifesp.br}
\address{Universidade Federal de São Paulo, Departamento de Ciências Exatas
e da Terra, Rua Prof. Artur Riedel, 275, 09972-270, Diadema, SP, Brazil}

\begin{abstract}
We propose a type of non-anticommutative superspace, with the interesting
property of relating to Lee-Wick type of higher derivatives theories,
which are known for their interesting properties, and have lead to
proposals of phenomenologicaly viable higher derivatives extensions
of the Standard Model. The deformation of superspace we consider does
not preserve supersymmetry or associativity in general; however, we
show that a non-anticommutative version of the Wess-Zumino model can
be properly defined. In fact, the definition of chiral and antichiral
superfields turns out to be simpler in our case than in the well known
${\cal N}=1/2$ supersymmetric case. We show that, when the theory
is truncated at the first nontrivial order in the deformation parameter,
supersymmetry is restored, and we end up with a well known Lee-Wick
type of higher derivative extension of the Wess-Zumino model. Thus
we show how non-anticommutative could provide an alternative mechanism
for generation of these kind of higher derivative theories.
\end{abstract}

\maketitle

\section{Introduction}

The fundamental nature of spacetime at the Planck scale is still considered
to be an open question, and the emergence of some type of noncommutativity
between coordinates has been studied in different contexts for some
time\,\cite{snyder:1946qz,lukierski:1992dt,majid:1994cy,doplicher:1994tu,seiberg:1999vs,chaichian:2000si}.
One of the nontrivial aspects of these studies is the fate of spacetime
symmetries when noncommutativity is considered: the Poincaré symmetry
can be broken by the noncommutativity such as in the low energy description
of some string dynamics with a fixed background\,\cite{seiberg:1999vs},
it can be preserved in some adequate sense\,\cite{chaichian:2004yh,balachandran:2007vx},
or deformed in an algebraically consistent way, such as in the context
of kappa deformed spacetimes\,\cite{majid:1994cy,lukierski:1992dt}.
Poincaré symmetry is one of the conceptual cornerstones of the current
understanding of elementary interactions, and besides that, in the
last decades the possibility of small deviations of Lorentz symmetry
has been studied in a systematic way\,\cite{colladay:1998fq}, providing
increasingly stronger phenomenological constraints on Lorentz violation\,\cite{Kostelecky:2008ts}:
these facts reinforce the interest in studying the fundamental symmetries
of proposed deformations of spacetime, including those involving noncommuting
coordinates.

Supersymmetry is known to be the essentially unique consistent extension
of the Poincaré symmetry\,\cite{Haag:1974qh}, and not surprisingly,
the possibility of implementing noncommutative deformations of superspace
has also been studied for some time. Following the discovery of the
connection between noncommutative geometry and superstring theory,
a string configuration where a supersymmetric field theory with spacetime
noncommutativity appears as a low energy limit was identified in\,\cite{Chu:1999ij}.
In this case, supersymmetry is left untouched by the deformation,
which affects only the spacetime coordinates $x^{\mu}$, while fermionic
coordinates $\theta^{\alpha}$, $\bar{\theta}^{\alpha}$ satisfy usual
anticommutation relations: because of this, all the superspace formalism
for studying classical and quantum aspects of supersymmetric field
theories can be preserved in the study of its noncommutative counterparts\,\cite{Ferrara:2000mm,Terashima:2000xq,Zanon:2000gy,ferrari:2004ex}. 

In\,\cite{klemm:2001yu}, a general discussion of non trivial (anti)commutations
relations between spacetime (and spinorial) coordinates of superspace
was presented, and not surprisingly the discussion becomes more complicated
if the algebra of the fermionic coordinates $\theta^{\alpha}$, $\bar{\theta}^{\dot{\alpha}}$
is modified: it was pointed out that in the non-anticommutative case,
in general the supersymmetry algebra is deformed, and the associativity
of the products of superfields is lost. The authors of\,\cite{klemm:2001yu}
discussed a special case where associativity is maintained, and a
more general study of non-anticommutative deformations of superspace
which preserve the essential properties of supersymmetry and associativity
can be developed using the formalism of twist deformations of Hopf
algebras\,\cite{kobayashi:2004ep,Ihl:2005zd,Irisawa:2006xx}.

In\,\cite{seiberg:2003yz}, another interesting connection between
string theory and non-anticommutative theory was unveiled. In this
case the anticommutation relation involving fermionic coordinates
$\theta^{\alpha}$ is the primary source of deformation, and the algebra
of half of the supersymmetry generators is broken, hence the name
${\cal N}=1/2$ supersymmetry. The notion of chirality has to be dealt
with properly, and a consistent interacting theory involving gauge
and scalar superfields have been defined. The spurion technique can
be used to treat these theories at the quantum level, and their renormalizability
have been studied\,\cite{romagnoni:2003xt,Grisaru:2003fd,Grisaru:2004qw,Grisaru:2005we}.
We will review some aspects of the ${\cal N}=1/2$ supersymmetric
field theories in Sec.\,\ref{sec:Non(anti)commutative-Superspace},
since they will be relevant to our work.

In this work, we propose a new type of deformed superspace, one in
which the anticommutator involving $\theta^{\alpha}$ and $\bar{\theta}^{\dot{\alpha}}$
is deformed. We argue that this possibility has very interesting properties,
mainly that at the second order in the deformation parameter, the
scalar superfield theory defined in this non-anticommutative superspace
exactly coincides with the higher derivative supersymmetric Wess-Zumino
model studied, for example, in\,\cite{Antoniadis:2007xc}. This is
interesting because Lee-Wick type of theories\,\cite{Lee:1969fy,Lee:1970iw}
have recently been revived in higher derivative extensions of the
Standard Model, which attempts to be phenomenologically viable and
free of hierarchy problems\,\cite{Grinstein:2007mp,Carone:2008bs,Figy:2011yu,Carone:2014kla}.
Supersymmetric models with higher derivative operators have also attracted
some attention at the formal level in recent years\,\cite{Dias:2012fi,Gama:2013rsa,Gama:2014fca}.
The connection found in this work, between a particular form of deformation
of superspace and the higher derivative supersymmetric model studied
in\,\cite{Antoniadis:2007xc,Dias:2012fi} is even more interesting
because the former breaks supersymmetry in general, however we will
show that at the second order of the deformation parameter, supersymmetry
is fully restored in our model.

This work is organized as follows. In Sec.\,\ref{sec:Non(anti)commutative-Superspace}
we review some aspects of the ${\cal N}=1/2$ supersymmetry construction,
and propose our deformed superspace, discussing its main properties.
The supersymmetric algebra in this deformed superspace is studied
in Sec.\,\ref{sec:Operators-Algebra}, and it is shown to be also
deformed, signaling that supersymmetry is not preserved by our construction
in general. In Sec.\,\ref{sec:Action-in-the}, however, we show that
we can define an action for scalar superfields in this deformed superspace,
and how it becomes equivalent, at the second order of the deformation
parameter, to a higher derivative, supersymmetric invariant Wess-Zumino
model. Sec.\,\ref{sec:Concluding-Remarks} contains our conclusions
and perspectives.

\section{\label{sec:Non(anti)commutative-Superspace}Deformed Superspace}

We start by reviewing some of the basic aspects of the non-anticommutative
superspace discussed in\,\cite{seiberg:2003yz}. 

The four dimensional superspace has bosonic coordinates $x^{\mu}$
which are spacetime vectors, and fermionic coordinates $\theta^{\alpha}$,
$\bar{\theta}^{\dot{\alpha}}$ which are two component Weyl spinors,
whose indices can be lowered and raised with the antisymmetric symbols
$\epsilon^{\alpha\beta}$ and $\epsilon_{\alpha\beta}$, normalized
according to $\epsilon^{12}=-\epsilon_{12}=1$, and similarly for
dotted indices. It is also useful to introduce the symbols\begin{subequations}
\begin{align}
\left(\sigma^{\mu}\right)_{\alpha\dot{\alpha}} & =\left(1,\sigma_{1},\sigma_{2},\sigma_{3}\right)\thinspace,\\
\left(\bar{\sigma}^{\mu}\right)^{\dot{\alpha}\alpha} & =\left(1,-\sigma_{1},-\sigma_{2},-\sigma_{3}\right)\thinspace,\\
\left(\sigma^{\mu\nu}\right)_{\alpha}^{\thinspace\thinspace\beta} & =\left(\sigma^{\mu}\bar{\sigma}^{\nu}-\sigma^{\nu}\bar{\sigma}^{\mu}\right)_{\alpha}^{\thinspace\thinspace\beta}\thinspace,
\end{align}
\end{subequations}$\sigma_{i}$ being the Pauli matrices. Finally,
supersymmetry generators and supercovariant derivatives are defined
by\begin{subequations}\label{eq:QDs}
\begin{align}
Q_{\alpha}=\partial_{\alpha}-i\sigma_{\alpha\dot{\alpha}}^{\mu}\bar{\theta}^{\dot{\alpha}}\partial_{\mu}\  & ,\ \bar{Q}_{\dot{\alpha}}=-\bar{\partial}_{\dot{\alpha}}+i\theta^{\alpha}\sigma_{\alpha\dot{\alpha}}^{\mu}\partial_{\mu}\ ,\label{eq:Qs}\\
D_{\alpha}=\partial_{\alpha}+i\sigma_{\alpha\dot{\alpha}}^{\mu}\bar{\theta}^{\dot{\alpha}}\partial_{\mu}\  & ,\ \bar{D}_{\dot{\alpha}}=-\bar{\partial}_{\dot{\alpha}}-i\theta^{\alpha}\sigma_{\alpha\dot{\alpha}}^{\mu}\partial_{\mu}\ .\label{eq:Ds}
\end{align}
\end{subequations}We refer the reader to\,\cite{WessBagger} for
other notations and definitions that are used in this work. 

The deformation in superspace is introduced by assuming that the coordinates
$\theta^{\alpha}$ no longer anticommute, instead they satisfy
\begin{equation}
\left\{ \theta^{\alpha},\theta^{\beta}\right\} =c^{\alpha\beta}\thinspace,\label{eq:1}
\end{equation}
where $c^{\alpha\beta}$ is a two-dimensional symmetric matrix that
plays the role of a deformation parameter. The $\bar{\theta}^{\dot{\alpha}}$
coordinates anticommute with themselves and $\theta^{\alpha}$ as
usual, but commutation relations involving $x^{\mu}$ and $\theta^{\alpha}$
are changed to
\begin{align}
\left[x^{\mu},x^{\nu}\right] & =\overline{\text{\ensuremath{\theta\theta}}}c^{\alpha\beta}\epsilon_{\beta\gamma}\left(\sigma^{\mu\nu}\right)_{\alpha}^{\thinspace\thinspace\gamma}\,,\\
\left[x^{\mu},\theta^{\alpha}\right] & =ic^{\alpha\beta}\sigma_{\beta\dot{\alpha}}^{\mu}\bar{\theta}^{\dot{\alpha}}\thinspace.
\end{align}
This particular algebra is interesting because if written in terms
of chiral coordinates,
\begin{equation}
y^{\mu}=x^{\mu}+i\theta\sigma^{\mu}\bar{\theta}\thinspace,\label{eq:chiralcoords}
\end{equation}
it implies that 
\begin{equation}
\left[y^{\mu},y^{\nu}\right]=\left[y^{\mu},\theta^{\alpha}\right]=\left[y^{\mu},\bar{\theta}^{\dot{\alpha}}\right]=0\thinspace,\label{eq:2}
\end{equation}
which simplifies the definition of chiral superfields. 

Due to Eq.\,\eqref{eq:1}, products of the $\theta$ coordinate should
be properly ordered, and when multiplying two functions of $\theta$,
the result should be reordered; this can be accomplished by means
of a star product,
\begin{equation}
f\left(\theta\right)\star g\left(\theta\right)=f\left(\theta\right)\exp\left(-\frac{c^{\alpha\beta}}{2}\thinspace\overleftarrow{\frac{\partial}{\partial\theta^{\alpha}}}\thinspace\overrightarrow{\frac{\partial}{\partial\theta^{\alpha}}}\right)g\left(\theta\right)\thinspace,\label{eq:3}
\end{equation}
which implements the Weyl (symmetric) ordering. It is noteworthy that
the exponential in Eq.\,\eqref{eq:3} is finite due to the anticommuting
nature of the derivatives $\partial/\partial\theta^{\alpha}$. Chiral
superfields are defined to be functions of $y$ and $\theta$ alone,
and when multiplying two chiral superfields via the star product,
the result is still a chiral superfield. 

Finally, when studying the anticommutation relations between the supersymmetry
generators $Q_{\alpha}$ and $\bar{Q}_{\dot{\alpha}}$, one finds
the standard relations
\begin{align}
\left\{ Q_{\alpha},\bar{Q}_{\dot{\alpha}}\right\}  & =2i\sigma_{\alpha\dot{\alpha}}^{\mu}\frac{\partial}{\partial y^{\mu}}\thinspace,\\
\left\{ Q_{\alpha},Q_{\beta}\right\}  & =0\thinspace,
\end{align}
on the other hand,
\begin{equation}
\left\{ \bar{Q}_{\dot{\alpha}},\bar{Q}_{\dot{\beta}}\right\} =-4c^{\alpha\beta}\sigma_{\alpha\dot{\alpha}}^{\mu}\sigma_{\beta\dot{\beta}}^{\nu}\thinspace\frac{\partial^{2}}{\partial y^{\mu}\partial y^{\nu}}\thinspace,
\end{equation}
signaling that \emph{half} of the supersymmetry generators is broken,
thus the name ${\cal N}=1/2$ supersymmetry for this construction.

One can continue to define antichiral and vector superfields, as done
in\,\cite{seiberg:2003yz}, but at this point we shall present our
proposal for a different non-anticommutative superspace. Our proposal
is defined via a different star product that will be responsible for
reordering functions of $\theta$ and $\bar{\theta}$ that are multiplied
together, 
\begin{align}
f(z)\star g(z) & =f(z)\exp\left[\frac{\xi}{2}C^{\alpha\dot{\alpha}}\left(\stackrel{\leftarrow}{D}_{\alpha}\stackrel{\rightarrow}{\bar{D}}_{\dot{\alpha}}+\stackrel{\leftarrow}{\bar{D}}_{\dot{\alpha}}\stackrel{\rightarrow}{D}_{\alpha}\right)\right]g(z)\nonumber \\
 & =fg+\frac{\xi}{2}(-1)^{s_{f}}\left[\left(D_{\alpha}f\right)\left(\bar{D}_{\dot{\alpha}}g\right)+\left(\bar{D}_{\dot{\alpha}}f\right)\left(D_{\alpha}g\right)\right]\nonumber \\
 & -\frac{\xi^{2}}{16}|C|\left[\left(D^{2}f\right)\left(\bar{D}^{2}g\right)+\left(\bar{D}^{2}f\right)\left(D^{2}g\right)\right]\nonumber \\
 & -\frac{\xi^{2}}{8}C^{\alpha\dot{\alpha}}C^{\beta\dot{\beta}}\left[(\bar{D}_{\dot{\beta}}D_{\alpha}f)(\bar{D}_{\dot{\alpha}}D_{\beta}g)\right.\nonumber \\
 & \left.\hspace{1cm}+(D_{\beta}\bar{D}_{\dot{\alpha}}f)(D_{\alpha}\bar{D}_{\dot{\beta}}g)\right]+\mathcal{O}(\xi^{3})\ ,\label{Moyal}
\end{align}
where $C^{\alpha\dot{\alpha}}$ is a symmetric tensor, $|C|=\frac{1}{2}C^{\alpha\dot{\alpha}}C^{\beta\dot{\beta}}\epsilon_{\alpha\beta}\epsilon_{\dot{\alpha}\dot{\beta}}$
is the determinant of $C^{\alpha\dot{\alpha}}$, 
\begin{equation}
\xi=\frac{1}{M^{2}}\thinspace,\label{eq:xi}
\end{equation}
$M$ being a very large mass scale, whose significance will become
clear afterwards, and $s_{f}$ is the parity of the function $f(z)$,
i.e. $s_{f}=0$ if $f$ is bosonic and $s_{f}=1$ if $f$ is fermionic. 

Based on the definition in Eq.\,\eqref{Moyal}, we can calculate
the deformed anticommutator between the fermionic variables $\theta$
and $\bar{\theta}$, obtaining 
\begin{align}
\left\{ \theta^{\alpha},\bar{\theta}^{\dot{\alpha}}\right\} _{\star} & =\theta^{\alpha}\star\bar{\theta}^{\dot{\alpha}}+\bar{\theta}^{\dot{\alpha}}\star\theta^{\alpha}\nonumber \\
 & =\theta^{\alpha}\bar{\theta}^{\dot{\alpha}}+\frac{\xi}{2}C^{\beta\dot{\beta}}(-1)^{1}(D_{\beta}\theta^{\alpha})(\bar{D}_{\dot{\beta}}\bar{\theta}^{\dot{\alpha}})\nonumber \\
 & +\bar{\theta}^{\dot{\alpha}}\theta^{\alpha}+\frac{\xi}{2}C^{\beta\dot{\beta}}(-1)^{1}(\bar{D}_{\dot{\beta}}\bar{\theta}^{\dot{\alpha}})(D_{\beta}\theta^{\alpha})\nonumber \\
 & =\xi C^{\alpha\dot{\alpha}}\thinspace.\label{Moyalthetas}
\end{align}
In the same fashion, we proceed to calculate the complete algebra
of the variables $x^{\mu}$, $\theta^{\alpha}$, $\bar{\theta}^{\dot{\alpha}}$
and also the chiral coordinates $y^{\mu}$ (see Eq.\,\eqref{eq:chiralcoords})
and antichiral ones,
\begin{equation}
\bar{y}^{\mu}=x^{\mu}-i\theta\sigma^{\mu}\bar{\theta}\thinspace.\label{eq:antichiral}
\end{equation}
The result follows:\begin{subequations}\label{eq:anticommutation}
\begin{align}
\left\{ \theta^{\alpha},\bar{\theta}^{\dot{\alpha}}\right\} _{\star} & =\xi C^{\alpha\dot{\alpha}}\ ,\\
\left\{ \theta^{\alpha},\theta^{\beta}\right\} _{\star} & =\left\{ \bar{\theta}^{\dot{\alpha}},\bar{\theta}^{\dot{\beta}}\right\} _{\star}=0\ ,
\end{align}
\end{subequations}\begin{subequations}
\begin{align}
\left[x^{\mu},x^{\nu}\right]_{\star} & =\xi C^{\alpha\dot{\beta}}(\sigma_{\alpha\dot{\alpha}}^{\mu}\sigma_{\beta\dot{\beta}}^{\nu}-\sigma_{\alpha\dot{\alpha}}^{\nu}\sigma_{\beta\dot{\beta}}^{\mu})\bar{\theta}^{\dot{\alpha}}\theta^{\beta}\ ,\\
\left[x^{\mu},\theta^{\alpha}\right]_{\star} & =-i\xi C^{\alpha\dot{\beta}}\sigma_{\beta\dot{\beta}}^{\mu}\theta^{\beta}\ ,\\
\left[x^{\mu},\bar{\theta}^{\dot{\alpha}}\right]_{\star} & =-i\xi C^{\beta\dot{\alpha}}\sigma_{\beta\dot{\beta}}^{\mu}\bar{\theta}^{\dot{\beta}}\ ,
\end{align}
\end{subequations}\begin{subequations}
\begin{align}
\left[x^{\mu},y^{\nu}\right]_{\star} & =-2\xi C^{\beta\dot{\alpha}}\sigma_{\alpha\dot{\alpha}}^{\mu}\sigma_{\beta\dot{\beta}}^{\nu}\bar{\theta}^{\dot{\beta}}\theta^{\alpha}\ ,\\
\left[x^{\mu},\bar{y}^{\nu}\right]_{\star} & =2\xi C^{\alpha\dot{\beta}}\sigma_{\alpha\dot{\alpha}}^{\mu}\sigma_{\beta\dot{\beta}}^{\nu}\bar{\theta}^{\dot{\alpha}}\theta^{\beta}\ ,\\
\left[y^{\mu},y^{\nu}\right]_{\star} & =\left[\bar{y}^{\mu},\bar{y}^{\nu}\right]_{\star}=0\ ,\label{eq:4}\\
\left[y^{\mu},\bar{y}^{\nu}\right]_{\star} & =4\xi C^{\alpha\dot{\beta}}\sigma_{\alpha\dot{\alpha}}^{\mu}\sigma_{\beta\dot{\beta}}^{\nu}\bar{\theta}^{\dot{\alpha}}\theta^{\beta}\ ,
\end{align}
\end{subequations}and\begin{subequations} 
\begin{align}
\left[y^{\mu},\theta^{\alpha}\right]_{\star} & =0\ ,\label{eq:5}\\
\left[y^{\mu},\bar{\theta}^{\dot{\alpha}}\right]_{\star} & =-2i\xi C^{\beta\dot{\alpha}}\sigma_{\beta\dot{\beta}}^{\mu}\bar{\theta}^{\dot{\beta}}=2\left[x^{\mu},\bar{\theta}^{\dot{\alpha}}\right]_{\star}\ ,\\
\left[\bar{y}^{\mu},\theta^{\alpha}\right]_{\star} & =-2i\xi C^{\alpha\dot{\beta}}\sigma_{\beta\dot{\beta}}^{\mu}\theta^{\beta}=2\left[x^{\mu},\theta^{\alpha}\right]_{\star}\ ,\\
\left[\bar{y}^{\mu},\bar{\theta}^{\dot{\alpha}}\right]_{\star} & =0\ .\label{eq:6}
\end{align}
\end{subequations}

Eqs.\,\eqref{eq:anticommutation} define the essential non-anticommutative
properties of this superspace. Most interesting are Eqs.\,\eqref{eq:4},\,\eqref{eq:5}
and\,\eqref{eq:6}, which show that the commutation relations involving
chiral coordinates $y^{\mu}$ and $\theta^{\alpha}$, as well as those
involving antichiral coordinates $\bar{y}^{\mu}$ and $\bar{\theta}^{\dot{\alpha}}$,
are not changed. This greatly simplifies the definition of chiral
and antichiral superfields. For example, chiral superfields are of
the form $\Phi\left(y^{\mu},\theta^{\alpha}\right)$, and since $y^{\mu}$
and $\theta^{\alpha}$ exhibit trivial (anti)commutation relations,
chiral superfields do not have to be reordered and, clearly, the (usual)
product of chiral superfields is again a chiral superfield. That means
the holomorphic potential $W\left(\Phi\right)$ of the Wess-Zumino
model is not affected by the non-anticommutativity. The same happens
for antichiral superfields and, consequently, for the antiholomorphic
potential. Any nontrivial modification brought by the superspace deformation
to our model will be in the Kähler part of the action, $\int d^{8}z\thinspace K\left(\Phi,\overline{\Phi}\right)$,
which, in the standard Wess-Zumino action at the classical level,
reduces to the kinetic term $\int d^{8}z\thinspace\Phi\overline{\Phi}$.

This last observation is also very important for the following: the
star product in Eq.\,\eqref{Moyal} is non associative (one can see
this as a corollary of the general discussion present in\,\cite{klemm:2001yu},
or see discussion at the end of the next section). Therefore, defining
star products of more than two superfields could introduce an ambiguity
in the definition of a deformed action. However, since the star product
does not appear in the potentials of the Wess-Zumino action, we can
define a deformed Wess-Zumino model in our deformed superspace without
dealing with this non associativity, at least at the classical level.

\section{\label{sec:Operators-Algebra}Operators Algebra}

We present in this section the algebra of the supersymmetric generators
and the covariant derivatives in our deformed superspace, which can
be calculated directly from the definitions contained in the last
section. 

The only non vanishing anticommutators are,\begin{subequations}\label{algebra}
\begin{align}
\left\{ Q_{\alpha},\bar{Q}_{\dot{\alpha}}\right\} _{\star} & =2i\sigma_{\alpha\dot{\alpha}}^{\mu}\partial_{\mu}+\xi C^{\beta\dot{\beta}}\sigma_{\alpha\dot{\beta}}^{\mu}\sigma_{\beta\dot{\alpha}}^{\nu}\partial_{\mu}\partial_{\nu}\ ,\\
\left\{ D_{\alpha},\bar{D}_{\dot{\alpha}}\right\} _{\star} & =-2i\sigma_{\alpha\dot{\alpha}}^{\mu}\partial_{\mu}+\xi C^{\beta\dot{\beta}}\sigma_{\alpha\dot{\beta}}^{\mu}\sigma_{\beta\dot{\alpha}}^{\nu}\partial_{\mu}\partial_{\nu}\ ,\\
\left\{ Q_{\alpha},\bar{D}_{\dot{\alpha}}\right\} _{\star} & =-\xi C^{\beta\dot{\beta}}\sigma_{\alpha\dot{\beta}}^{\mu}\sigma_{\beta\dot{\alpha}}^{\nu}\partial_{\mu}\partial_{\nu}\ ,\\
\left\{ \bar{Q}_{\dot{\alpha}},D_{\alpha}\right\} _{\star} & =-\xi C^{\beta\dot{\beta}}\sigma_{\alpha\dot{\beta}}^{\mu}\sigma_{\beta\dot{\alpha}}^{\nu}\partial_{\mu}\partial_{\nu}\ .
\end{align}
\end{subequations}

In a general way, defining $D_{A}=(D_{\alpha},\bar{D}_{\dot{\alpha}},\partial_{\mu})$
and $Q_{A}=(Q_{\alpha},\bar{Q}_{\dot{\alpha}},-\partial_{\mu})$,
the deformed algebra will have the following structure\,\cite{klemm:2001yu}
\begin{align}
\left[Q_{A},Q_{B}\right\} _{\star} & =T_{AB}^{\;\;\;\;\; C}Q_{C}+R_{AB}^{\;\;\;\;\; CD}Q_{C}Q_{D}\ ,\nonumber \\
\left[D_{A},D_{B}\right\} _{\star} & =T_{AB}^{\;\;\;\;\; C}D_{C}+R_{AB}^{\;\;\;\;\; CD}D_{C}D_{D}\ ,\nonumber \\
\left[Q_{A},D_{B}\right\} _{\star} & =R_{AB}^{\;\;\;\;\; CD}Q_{C}D_{D}\ ,\label{generalAlgebra}
\end{align}
where $T_{AB}^{\;\;\;\;\; C}$ is the torsion and $R_{AB}^{\;\;\;\;\; CD}=-\frac{1}{8}P^{MN}T_{M[A}^{\;\;\;\;\;\;\; C}T_{B)N}^{\;\;\;\;\;\;\; D}$
is the curvature tensor. In our case, the only non vanishing tensors
of these type are 
\begin{align}
T_{\alpha\dot{\alpha}}^{\;\;\;\;\;\mu} & =-2i\sigma_{\beta\dot{\beta}}^{\nu}\delta_{\alpha}^{\thinspace\thinspace\beta}\delta_{\dot{\alpha}}^{\thinspace\thinspace\dot{\beta}}\delta_{\nu}^{\mu}\ ,\\
R_{\alpha\dot{\alpha}}^{\;\;\;\;\;\mu\nu} & =-\frac{1}{8}\xi C^{\beta\dot{\beta}}(T_{\alpha\dot{\beta}}^{\;\;\;\;\;\mu}T_{\beta\dot{\alpha}}^{\;\;\;\;\;\nu}+T_{\beta\dot{\alpha}}^{\;\;\;\;\;\mu}T_{\alpha\dot{\beta}}^{\;\;\;\;\;\nu})\nonumber \\
 & =\frac{1}{2}\xi C^{\beta\dot{\beta}}(\sigma_{\alpha\dot{\beta}}^{\mu}\sigma_{\beta\dot{\alpha}}^{\nu}+\sigma_{\beta\dot{\alpha}}^{\mu}\sigma_{\alpha\dot{\beta}}^{\nu})\ ,
\end{align}
therefore, the algebra in Eq.\,(\ref{algebra}) can be written as
\begin{align}
\left\{ Q_{\alpha},Q_{\dot{\alpha}}\right\} _{\star} & =T_{\alpha\dot{\alpha}}^{\;\;\;\;\;\mu}Q_{\mu}+R_{\alpha\dot{\alpha}}^{\;\;\;\;\;\mu\nu}Q_{\mu}Q_{\nu}\ ,\\
\left\{ D_{\alpha},D_{\dot{\alpha}}\right\} _{\star} & =T_{\alpha\dot{\alpha}}^{\;\;\;\;\;\mu}D_{\mu}+R_{\alpha\dot{\alpha}}^{\;\;\;\;\;\mu\nu}D_{\mu}D_{\nu}\ ,\\
\left\{ Q_{\alpha},D_{\dot{\alpha}}\right\} _{\star} & =R_{\alpha\dot{\alpha}}^{\;\;\;\;\;\mu\nu}Q_{\mu}D_{\nu}\ .
\end{align}
From this result, we see that the supersymmetry algebra is broken
in general in the deformed superspace. This could be seen as a setback
in our construction, however, in the next section, we will see how
this problem can be solved, at least if the parameter $\xi$ is considered
so small that the theory can be truncated at a certain order in $\xi$.

It is worthwhile to mention that, similar to the three-dimensional
non-anticommutative superspace discussed in\,\cite{ferrari:2006hg},
one could define nonlinear generators $\widetilde{Q}_{\alpha}$, $\widetilde{\bar{Q}}_{\dot{\alpha}}$
and supercovariant derivatives $\widetilde{D}_{\alpha}$, $\widetilde{\bar{D}}_{\dot{\alpha}}$
that respect the usual supersymmetry algebra, in the form\begin{subequations}\label{eq:deformedgenerators}
\begin{align}
\widetilde{Q}_{\alpha} & =Q_{\alpha}+\frac{i}{2}\xi C^{\beta\dot{\beta}}\partial_{\beta}\partial_{\alpha\dot{\beta}}\thinspace\thinspace;\thinspace\thinspace\widetilde{\bar{Q}}_{\dot{\alpha}}=\bar{Q}_{\dot{\alpha}}-\frac{i}{2}\xi C^{\beta\dot{\beta}}\partial_{\dot{\beta}}\partial_{\beta\dot{\alpha}}\thinspace,\\
\widetilde{D}_{\alpha} & =D_{\alpha}-\frac{i}{2}\xi C^{\beta\dot{\beta}}\partial_{\beta}\partial_{\alpha\dot{\beta}}\thinspace\thinspace;\thinspace\thinspace\widetilde{\bar{D}}_{\dot{\alpha}}=\bar{D}_{\dot{\alpha}}+\frac{i}{2}\xi C^{\beta\dot{\beta}}\partial_{\dot{\beta}}\partial_{\beta\dot{\alpha}}\thinspace.
\end{align}
\end{subequations}These generators do not respect standard Leibniz
rule and, being nonlinear, they could be naturally incorporated in
the formalism of Hopf algebras. This observation raises the question
of whether the deformation we defined in Eq.\,\eqref{Moyal} could
be understood as a supersymmetric Drinfeld twist as studied, for example,
in\,\cite{kobayashi:2004ep,Ihl:2005zd}. However, the answer to this
question is negative. Indeed, explicit calculation shows that the
``twist element'' that would correspond to Eq.\,\eqref{Moyal} do
not satisfy the two-cocycle condition that has to be satisfied to
define a Drinfeld twist. This two-cocycle condition is what guarantees
the associativity of the star product, so this last observation is
another way to state the fact that the deformed product defined in
Eq.\,\eqref{Moyal} is not associative um general, as discussed at
the end of Sec.\,\ref{sec:Non(anti)commutative-Superspace}.

\section{\label{sec:Action-in-the}Wess-Zumino model in the deformed superspace}

To construct a Wess-Zumino action in the deformed superspace, we need
first to calculate the products between superfields with different
chiralities. Due to Eq.\,\eqref{Moyalthetas} they do not commute,
and should be multiplied using the star product\,\eqref{Moyal} to
properly reorder the non-anticommuting coordinates. The star products
between superfields $\Phi$ and $\bar{\Phi}$, truncated at the order
$\xi^{2}$ is given by, 
\begin{align}
\Phi\star\bar{\Phi} & =\Phi\bar{\Phi}+\frac{\xi}{2}C^{\alpha\dot{\alpha}}(D_{\alpha}\Phi)(\bar{D}_{\dot{\alpha}}\bar{\Phi})\nonumber \\
 & -\frac{\xi^{2}}{16}|C|(D^{2}\Phi)(\bar{D}^{2}\bar{\Phi})+\mathcal{O}(\xi^{3})\ ,\label{PhibarPhi}
\end{align}
\begin{align}
\bar{\Phi}\star\Phi & =\bar{\Phi}\Phi+\frac{\xi}{2}C^{\alpha\dot{\alpha}}(\bar{D}_{\dot{\alpha}}\bar{\Phi})(D_{\alpha}\Phi)\nonumber \\
 & -\frac{\xi^{2}}{16}|C|(\bar{D}^{2}\bar{\Phi})(D^{2}\Phi)+\mathcal{O}(\xi^{3})\ .\label{barPhiPhi}
\end{align}
These two expressions are different, so we have an ambiguity in the
generalization of the Wess-Zumino kinetic term $\int d^{8}z\thinspace\Phi\overline{\Phi}$
to the deformed superspace. We adopt a symmetric prescription, 
\begin{equation}
\Phi\star\bar{\Phi}+\bar{\Phi}\star\Phi=2\Phi\bar{\Phi}-\frac{\xi^{2}}{8}|C|(D^{2}\Phi)(\bar{D}^{2}\bar{\Phi})+\mathcal{O}(\xi^{3})\ ,\label{sumPhibarPhi}
\end{equation}
where in obtaining the right hand side we used the symmetry of $C^{\alpha\dot{\alpha}}$
and the fact that $D_{\alpha}\Phi\thinspace\bar{D}_{\dot{\alpha}}\bar{\Phi}=-\bar{D}_{\dot{\alpha}}\bar{\Phi}\thinspace D_{\alpha}\Phi$. 

We use Eq.\,\eqref{sumPhibarPhi} to define the free Wess-Zumino
model in the deformed superspace as 
\begin{eqnarray}
S & = & \frac{1}{2}\int d^{8}z(\Phi\star\bar{\Phi}+\bar{\Phi}\star\Phi)\nonumber \\
 & = & \int d^{8}z\left[\Phi\bar{\Phi}-\frac{\xi^{2}}{16}|C|(D^{2}\Phi)(\bar{D}^{2}\bar{\Phi})+\mathcal{O}(\xi^{3})\right]\ .\label{actionKahler}
\end{eqnarray}
Integrating by parts and using the relation $D^{2}\bar{D}^{2}\bar{\Phi}=16\Box\bar{\Phi}$,
we obtain, 
\begin{equation}
S=\int d^{8}z\left[\Phi(1-\xi^{2}|C|\Box)\bar{\Phi}+\mathcal{O}(\xi^{3})\right]\ .\label{actionBox}
\end{equation}
In terms of component fields, this last equation is written as 
\begin{align}
S & =\int d^{4}x\left[A(1-\xi^{2}|C|\Box)\Box\bar{A}+i\partial_{\mu}\psi\sigma^{\mu}(1-\xi^{2}|C|\Box)\bar{\psi}\right.\nonumber \\
 & \left.\hspace{1cm}+F(1-\xi^{2}|C|\Box)\bar{F}+\mathcal{O}(\xi^{3})\right]\ .\label{actionBoxcomponents}
\end{align}

Eqs.\,(\ref{actionBox}) and\,(\ref{actionBoxcomponents}), together
with the fact pointed out in Sec.\,\ref{sec:Non(anti)commutative-Superspace}
that the (anti)holomorphic potentials are not modified by the star
product, mean that the net effect of the superspace deformation is
the introduction of Lee-Wick type terms in the action. Most interestingly,
the action in Eq.\,\eqref{actionBox} is supersymmetric invariant.
This is a result of the symmetric prescription adopted in Eq.\,\eqref{sumPhibarPhi},
which cancels the linear terms present in Eqs.\,\eqref{PhibarPhi}
and\,\eqref{barPhiPhi}: these linear terms are the only non supersymmetric
terms up to order $\xi^{2}$. Indeed, the $\mathcal{O}(\xi^{1})$
term in Eq.\,\eqref{barPhiPhi}, for example, is written in terms
of component fields as follows,
\begin{align}
S_{\bar{D}\bar{\Phi}D\Phi} & =\int d^{4}x\,\frac{\xi}{2}\, C^{\alpha\dot{\alpha}}\left[i\,\sigma_{\alpha\dot{\alpha}}^{\mu}\left(\partial_{\mu}A\cdot\Box\bar{A}-\Box A\,\partial_{\mu}A+\bar{F}\partial_{\mu}F-F\partial_{\mu}\bar{F}\right)\right.+\nonumber \\
 & \left.+\frac{1}{2}\left(\psi_{\alpha}\Box\bar{\psi}_{\dot{\alpha}}+\Box\psi_{\alpha}\cdot\bar{\psi}_{\dot{\alpha}}\right)+\frac{1}{2}\zeta_{\alpha\beta\,\dot{\alpha}\dot{\beta}}^{\mu\nu}\left(\partial_{\mu}\psi^{\beta}\partial_{\nu}\bar{\psi}^{\dot{\beta}}+\partial_{\nu}\psi^{\beta}\cdot\partial_{\mu}\bar{\psi}^{\dot{\beta}}\right)\right],
\end{align}
where
\begin{equation}
\zeta_{\alpha\beta\,\dot{\alpha}\dot{\beta}}^{\mu\nu}=\sigma_{\alpha\dot{\alpha}}^{\mu}\sigma_{\beta\dot{\beta}}^{\nu}+\sigma_{\alpha\dot{\beta}}^{\nu}\sigma_{\beta\dot{\alpha}}^{\mu}\thinspace.
\end{equation}
By integrating by parts, $S_{\bar{D}\bar{\Phi}D\Phi}$ can be cast
as
\begin{align}
S_{\bar{D}\bar{\Phi}D\Phi} & =\int d^{4}x\,\frac{\xi}{2}\, C^{\alpha\dot{\alpha}}\left[i\,\sigma_{\alpha\dot{\alpha}}^{\mu}\left(2\,\partial_{\mu}\Box A\cdot\bar{A}+2\,\bar{F}\partial_{\mu}F\right)+\right.\nonumber \\
 & \left.+\left(\Box\psi_{\alpha}\cdot\bar{\psi}_{\dot{\alpha}}\right)-\zeta_{\alpha\beta\,\dot{\alpha}\dot{\beta}}^{\mu\nu}\,\partial_{\nu}\partial_{\mu}\psi^{\beta}\cdot\bar{\psi}^{\dot{\beta}}\right]\thinspace.\label{eq:linear}
\end{align}
It is a straightforward, yet cumbersome calculation, to verify that
Eq.\,\eqref{eq:linear} is not invariant under the standard supersymmetric
transformations.

Therefore, despite the broken supersymmetric algebra discussed in
Sec.\,\ref{sec:Operators-Algebra}, if the model is truncated at
order $\xi^{2}$, we can use the ordering ambiguity introduced by
the deformation itself to cancel the non symmetric parts, thus restoring
supersymmetry. Actually, we verified that the $\mathcal{O}(\xi^{3})$
terms in Eqs.\,\eqref{barPhiPhi} and\,\eqref{barPhiPhi} behave
in the very same way as the $\mathcal{O}(\xi^{1})$ terms, i.e., they
cancel with the symmetric prescription adopted in Eq.\,\eqref{sumPhibarPhi},
so actually Eq.\,\eqref{actionBox} holds up to $\mathcal{O}(\xi^{4})$
terms.

As a conclusion, we can state that, at the first nontrivial order
in the deformation parameter $\xi$, the particular form of the non-anticommutativity
considered by us presents itself as an alternative mechanism for the
generation of a very important class of higher derivative theories,
which usually are considered to appear as a result of the integration
of fields with very large mass\,\cite{Antoniadis:2007xc}.

Finally, from Eq.\,\eqref{actionBox} we see that the parameter $\xi$
is indeed inversely proportional to the square of the mass scale where
the higher dimensional operators are relevant (see discussion in\,\cite{Antoniadis:2007xc}
for example); this justifies our definition\,\eqref{eq:xi}.

\section{\label{sec:Concluding-Remarks}Conclusions and Perspectives}

In this work, we proposed an alternative deformed superspace, which
has the interesting property of being connected with higher derivative
theories that are well studied in the literature, relevant even to
phenomenological models based on the Standard Model\,\cite{Carone:2014kla}.
We defined a natural non-anticommutative version of the Wess-Zumino
model in this superspace, showing that, despite the supersymmetry
algebra being broken in general, supersymmetry is restored by a proper
ordering prescription, if the theory is truncated at the first non
trivial order in the deformation parameter $\xi$. These results leave
us with some open questions that deserve further investigation. The
first would be the extension of this work to gauge theories. Another
interesting perspective is to investigate quantum corrections and
study the consistency of the model at the quantum level.

\bigskip{}

\textbf{Acknowledgements.} This work was partially supported by the
Brazilian agencies Coordenação de Aperfeiçoamento de Pessoal de Nivel
Superior (CAPES), Conselho Nacional de Desenvolvimento Científico
e Tecnológico (CNPq) and Fundação de Amparo à Pesquisa do Estado de
São Paulo (FAPESP), via the following grants: CNPq 482874/2013-9 and
FAPESP 2013/22079-8 (AFF), CAPES PhD grant (CAP).


\begin{thebibliography}{10}
\expandafter\ifx\csname url\endcsname\relax
  \def\url#1{\texttt{#1}}\fi
\expandafter\ifx\csname urlprefix\endcsname\relax\def\urlprefix{URL }\fi
\expandafter\ifx\csname href\endcsname\relax
  \def\href#1#2{#2} \def\path#1{#1}\fi

\bibitem{snyder:1946qz}
H.~S. Snyder, {Quantized space-time}, Phys. Rev. 71 (1947) 38--41.

\bibitem{lukierski:1992dt}
J.~Lukierski, A.~Nowicki, H.~Ruegg, {New quantum Poincare algebra and k
  deformed field theory}, Phys. Lett. B293 (1992) 344--352.

\bibitem{majid:1994cy}
S.~Majid, H.~Ruegg, {Bicrossproduct structure of kappa Poincare group and
  noncommutative geometry}, Phys. Lett. B334 (1994) 348--354.
\newblock \href {http://arxiv.org/abs/hep-th/9405107}
  {\path{arXiv:hep-th/9405107}}.

\bibitem{doplicher:1994tu}
S.~Doplicher, K.~Fredenhagen, J.~E. Roberts, {The Quantum structure of
  space-time at the Planck scale and quantum fields}, Commun. Math. Phys. 172
  (1995) 187--220.
\newblock \href {http://arxiv.org/abs/hep-th/0303037}
  {\path{arXiv:hep-th/0303037}}.

\bibitem{seiberg:1999vs}
N.~Seiberg, E.~Witten, {String theory and noncommutative geometry}, JHEP 09
  (1999) 032.
\newblock \href {http://arxiv.org/abs/hep-th/9908142}
  {\path{arXiv:hep-th/9908142}}.

\bibitem{chaichian:2000si}
M.~Chaichian, M.~M. Sheikh-Jabbari, A.~Tureanu, {Hydrogen atom spectrum and the
  Lamb shift in noncommutative {QED}}, Phys. Rev. Lett. 86 (2001) 2716.
\newblock \href {http://arxiv.org/abs/hep-th/0010175}
  {\path{arXiv:hep-th/0010175}}.

\bibitem{chaichian:2004yh}
M.~Chaichian, P.~Presnajder, A.~Tureanu, {New concept of relativistic
  invariance in {NC} space-time: Twisted Poincare symmetry and its
  implications}, Phys. Rev. Lett. 94 (2005) 151602.
\newblock \href {http://arxiv.org/abs/hep-th/0409096}
  {\path{arXiv:hep-th/0409096}}.

\bibitem{balachandran:2007vx}
A.~P. Balachandran, A.~Pinzul, B.~A. Qureshi, {Twisted Poincar{\'e} Invariant
  Quantum Field Theories}, Phys. Rev. D77 (2008) 025021.
\newblock \href {http://arxiv.org/abs/0708.1779} {\path{arXiv:0708.1779}},
  \href {http://dx.doi.org/10.1103/PhysRevD.77.025021}
  {\path{doi:10.1103/PhysRevD.77.025021}}.

\bibitem{colladay:1998fq}
D.~Colladay, V.~A. Kostelecky, {Lorentz-violating extension of the standard
  model}, Phys. Rev. D58 (1998) 116002.
\newblock \href {http://arxiv.org/abs/hep-ph/9809521}
  {\path{arXiv:hep-ph/9809521}}.

\bibitem{Kostelecky:2008ts}
V.~A. Kostelecky, N.~Russell, {Data Tables for Lorentz and CPT Violation}, Rev.
  Mod. Phys. 83 (2011) 11.
\newblock \href {http://arxiv.org/abs/0801.0287} {\path{arXiv:0801.0287}},
  \href {http://dx.doi.org/10.1103/RevModPhys.83.11}
  {\path{doi:10.1103/RevModPhys.83.11}}.

\bibitem{Haag:1974qh}
R.~Haag, J.~T. Lopuszanski, M.~Sohnius, {All Possible Generators of
  Supersymmetries of the s Matrix}, Nucl.Phys. B88 (1975) 257.
\newblock \href {http://dx.doi.org/10.1016/0550-3213(75)90279-5}
  {\path{doi:10.1016/0550-3213(75)90279-5}}.

\bibitem{Chu:1999ij}
C.-S. Chu, F.~Zamora, {Manifest supersymmetry in noncommutative geometry}, JHEP
  0002 (2000) 022.
\newblock \href {http://arxiv.org/abs/hep-th/9912153}
  {\path{arXiv:hep-th/9912153}}, \href
  {http://dx.doi.org/10.1088/1126-6708/2000/02/022}
  {\path{doi:10.1088/1126-6708/2000/02/022}}.

\bibitem{Ferrara:2000mm}
S.~Ferrara, M.~Lledo, {Some aspects of deformations of supersymmetric field
  theories}, JHEP 0005 (2000) 008.
\newblock \href {http://arxiv.org/abs/hep-th/0002084}
  {\path{arXiv:hep-th/0002084}}, \href
  {http://dx.doi.org/10.1088/1126-6708/2000/05/008}
  {\path{doi:10.1088/1126-6708/2000/05/008}}.

\bibitem{Terashima:2000xq}
S.~Terashima, {A Note on superfields and noncommutative geometry}, Phys.Lett.
  B482 (2000) 276--282.
\newblock \href {http://arxiv.org/abs/hep-th/0002119}
  {\path{arXiv:hep-th/0002119}}, \href
  {http://dx.doi.org/10.1016/S0370-2693(00)00486-X}
  {\path{doi:10.1016/S0370-2693(00)00486-X}}.

\bibitem{Zanon:2000gy}
D.~Zanon, {Noncommutative perturbation in superspace}, Phys.Lett. B504 (2001)
  101--108.
\newblock \href {http://arxiv.org/abs/hep-th/0009196}
  {\path{arXiv:hep-th/0009196}}, \href
  {http://dx.doi.org/10.1016/S0370-2693(01)00271-4}
  {\path{doi:10.1016/S0370-2693(01)00271-4}}.

\bibitem{ferrari:2004ex}
A.~F. Ferrari, H.~O. Girotti, M.~Gomes, A.~Y. Petrov, A.~A. Ribeiro, V.~O.
  Rivelles, A.~J. da~Silva, {Towards a consistent noncommutative supersymmetric
  Yang- Mills theory: Superfield covariant analysis}, Phys. Rev. D70 (2004)
  085012.
\newblock \href {http://arxiv.org/abs/hep-th/0407040}
  {\path{arXiv:hep-th/0407040}}.

\bibitem{klemm:2001yu}
D.~Klemm, S.~Penati, L.~Tamassia, {Non(anti)commutative superspace}, Class.
  Quant. Grav. 20 (2003) 2905--2916.
\newblock \href {http://arxiv.org/abs/hep-th/0104190}
  {\path{arXiv:hep-th/0104190}}.

\bibitem{kobayashi:2004ep}
Y.~Kobayashi, S.~Sasaki, {Lorentz invariant and supersymmetric interpretation
  of noncommutative quantum field theory}, Int. J. Mod. Phys. A20 (2005)
  7175--7188.
\newblock \href {http://arxiv.org/abs/hep-th/0410164}
  {\path{arXiv:hep-th/0410164}}.

\bibitem{Ihl:2005zd}
M.~Ihl, C.~Saemann, {Drinfeld-twisted supersymmetry and non-anticommutative
  superspace}, JHEP 0601 (2006) 065.
\newblock \href {http://arxiv.org/abs/hep-th/0506057}
  {\path{arXiv:hep-th/0506057}}, \href
  {http://dx.doi.org/10.1088/1126-6708/2006/01/065}
  {\path{doi:10.1088/1126-6708/2006/01/065}}.

\bibitem{Irisawa:2006xx}
M.~Irisawa, Y.~Kobayashi, S.~Sasaki, {Drinfel'd twisted superconformal algebra
  and structure of unbroken symmetries}, Prog.Theor.Phys. 118 (2007) 83--96.
\newblock \href {http://arxiv.org/abs/hep-th/0606207}
  {\path{arXiv:hep-th/0606207}}, \href {http://dx.doi.org/10.1143/PTP.118.83}
  {\path{doi:10.1143/PTP.118.83}}.

\bibitem{seiberg:2003yz}
N.~Seiberg, {Noncommutative superspace, {N = 1/2} supersymmetry, field theory
  and string theory}, JHEP 06 (2003) 010.
\newblock \href {http://arxiv.org/abs/hep-th/0305248}
  {\path{arXiv:hep-th/0305248}}.

\bibitem{romagnoni:2003xt}
A.~Romagnoni, {Renormalizability of {N} = 1/2 Wess-Zumino model in superspace},
  JHEP 10 (2003) 016.
\newblock \href {http://arxiv.org/abs/hep-th/0307209}
  {\path{arXiv:hep-th/0307209}}.

\bibitem{Grisaru:2003fd}
M.~T. Grisaru, S.~Penati, A.~Romagnoni, {Two loop renormalization for
  nonanticommutative N = 1/2 supersymmetric WZ model}, JHEP 0308 (2003) 003.
\newblock \href {http://arxiv.org/abs/hep-th/0307099}
  {\path{arXiv:hep-th/0307099}}, \href
  {http://dx.doi.org/10.1088/1126-6708/2003/08/003}
  {\path{doi:10.1088/1126-6708/2003/08/003}}.

\bibitem{Grisaru:2004qw}
M.~T. Grisaru, S.~Penati, A.~Romagnoni, {Nonanticommutative superspace and N =
  1/2 WZ model}, Class.Quant.Grav. 21 (2004) S1391--1398.
\newblock \href {http://arxiv.org/abs/hep-th/0401174}
  {\path{arXiv:hep-th/0401174}}, \href
  {http://dx.doi.org/10.1088/0264-9381/21/10/012}
  {\path{doi:10.1088/0264-9381/21/10/012}}.

\bibitem{Grisaru:2005we}
M.~T. Grisaru, S.~Penati, A.~Romagnoni, {Non(anti)commutative sym theory:
  Renormalization in superspace}, JHEP 0602 (2006) 043.
\newblock \href {http://arxiv.org/abs/hep-th/0510175}
  {\path{arXiv:hep-th/0510175}}, \href
  {http://dx.doi.org/10.1088/1126-6708/2006/02/043}
  {\path{doi:10.1088/1126-6708/2006/02/043}}.

\bibitem{Antoniadis:2007xc}
I.~Antoniadis, E.~Dudas, D.~Ghilencea, {Supersymmetric Models with Higher
  Dimensional Operators}, JHEP 0803 (2008) 045.
\newblock \href {http://arxiv.org/abs/0708.0383} {\path{arXiv:0708.0383}},
  \href {http://dx.doi.org/10.1088/1126-6708/2008/03/045}
  {\path{doi:10.1088/1126-6708/2008/03/045}}.

\bibitem{Lee:1969fy}
T.~Lee, G.~Wick, {Negative Metric and the Unitarity of the S Matrix},
  Nucl.Phys. B9 (1969) 209--243.
\newblock \href {http://dx.doi.org/10.1016/0550-3213(69)90098-4}
  {\path{doi:10.1016/0550-3213(69)90098-4}}.

\bibitem{Lee:1970iw}
T.~Lee, G.~Wick, {Finite Theory of Quantum Electrodynamics}, Phys.Rev. D2
  (1970) 1033--1048.
\newblock \href {http://dx.doi.org/10.1103/PhysRevD.2.1033}
  {\path{doi:10.1103/PhysRevD.2.1033}}.

\bibitem{Grinstein:2007mp}
B.~Grinstein, D.~O'Connell, M.~B. Wise, {The Lee-Wick standard model},
  Phys.Rev. D77 (2008) 025012.
\newblock \href {http://arxiv.org/abs/0704.1845} {\path{arXiv:0704.1845}},
  \href {http://dx.doi.org/10.1103/PhysRevD.77.025012}
  {\path{doi:10.1103/PhysRevD.77.025012}}.

\bibitem{Carone:2008bs}
C.~D. Carone, R.~F. Lebed, {Minimal Lee-Wick Extension of the Standard Model},
  Phys.Lett. B668 (2008) 221--225.
\newblock \href {http://arxiv.org/abs/0806.4555} {\path{arXiv:0806.4555}},
  \href {http://dx.doi.org/10.1016/j.physletb.2008.08.050}
  {\path{doi:10.1016/j.physletb.2008.08.050}}.

\bibitem{Figy:2011yu}
T.~Figy, R.~Zwicky, {The other Higgses, at resonance, in the Lee-Wick extension
  of the Standard Model}, JHEP 1110 (2011) 145.
\newblock \href {http://arxiv.org/abs/1108.3765} {\path{arXiv:1108.3765}},
  \href {http://dx.doi.org/10.1007/JHEP10(2011)145}
  {\path{doi:10.1007/JHEP10(2011)145}}.

\bibitem{Carone:2014kla}
C.~D. Carone, R.~Ramos, M.~Sher, {LHC Constraints on the Lee-Wick Higgs
  Sector}, Phys.Lett. B732 (2014) 122--126.
\newblock \href {http://arxiv.org/abs/1403.0011} {\path{arXiv:1403.0011}},
  \href {http://dx.doi.org/10.1016/j.physletb.2014.03.025}
  {\path{doi:10.1016/j.physletb.2014.03.025}}.

\bibitem{Dias:2012fi}
M.~Dias, A.~Y. Petrov, J.~Senise, C.R., A.~da~Silva, {Effective potential for a
  SUSY Lee-Wick model: the Wess-Zumino case}\href
  {http://arxiv.org/abs/1212.5220} {\path{arXiv:1212.5220}}.

\bibitem{Gama:2013rsa}
F.~Gama, J.~Nascimento, A.~Y. Petrov, {Effective superpotential in the generic
  higher-derivative three-dimensional scalar superfield theory}, Phys.Rev.
  D88~(6) (2013) 065029.
\newblock \href {http://arxiv.org/abs/1308.5834} {\path{arXiv:1308.5834}},
  \href {http://dx.doi.org/10.1103/PhysRevD.88.065029}
  {\path{doi:10.1103/PhysRevD.88.065029}}.

\bibitem{Gama:2014fca}
F.~Gama, M.~Gomes, J.~Nascimento, A.~Y. Petrov, A.~da~Silva, {On the one-loop
  effective potential in the higher-derivative four-dimensional chiral
  superfield theory with a nonconventional kinetic term}, Phys.Lett. B733
  (2014) 247--252.
\newblock \href {http://arxiv.org/abs/1401.5414} {\path{arXiv:1401.5414}},
  \href {http://dx.doi.org/10.1016/j.physletb.2014.04.054}
  {\path{doi:10.1016/j.physletb.2014.04.054}}.

\bibitem{WessBagger}
J.~Wess, J.~Bagger, Supersymmetry and supergravity, Princeton, USA: Univ. Pr.,
  1992, 259 p.

\bibitem{ferrari:2006hg}
A.~F. Ferrari, M.~Gomes, J.~R. Nascimento, A.~Y. Petrov, A.~J. da~Silva, {The
  three-dimensional non-anticommutative superspace}, Phys. Rev. D74 (2006)
  125016.
\newblock \href {http://arxiv.org/abs/hep-th/0607087}
  {\path{arXiv:hep-th/0607087}}.

\end{thebibliography}
\end{document}